\documentclass[twocolumn,pra,aps]{revtex4}

\usepackage[dvipdfmx]{graphicx}
\usepackage{color, fancybox} 
\usepackage{amsfonts}
\usepackage{amssymb,amsmath,amsthm} 
\usepackage{braket}

\begin{document}
\newcommand{\mbold}[1]{\mbox{\boldmath $#1$}}
\newcommand{\sbold}[1]{\mbox{\boldmath ${\scriptstyle #1}$}}
\newcommand{\tr}{\,{\rm tr}\,}
\newcommand{\del}{\partial}
\newcommand{\mint}{\int\!\!}
\renewcommand{\arraystretch}{1.5}

\newtheorem{theorem}{Theorem}
\newtheorem*{theorem*}{Theorem}
\newtheorem{lemma}{Lemma}
\newtheorem*{lemma*}{Lemma}

\title{Gauge dependence of the Aharonov-Bohm phase in a quantum electrodynamics framework}
\author{A. Hayashi}
\email[E-mail: ]{hayashia@u-fukui.ac.jp}
\affiliation{Department of Applied Physics (Emeritus), 
          University of Fukui, Fukui 910-8507, Japan
}

\begin{abstract}
The Aharonov-Bohm (AB) phase is usually associated with a line integral of the electromagnetic 
vector potential generated by an external current source, such as a solenoid.  
According to this interpretation, the AB phase of a nonclosed path cannot be observed, as the integral 
depends on the gauge choice of the vector potential. 
Recent attempts to explain the AB effect through the interaction between a charged particle and an 
external current, mediated by the exchange of quantum photons, have assumed that the AB 
phase shift is proportional to the change in interaction energy between the charged particle and the 
external current source.  
As a result, these attempts argue that the AB phase change along 
a path does not depend on the gauge choice, and that the AB phase shift for a nonclosed path is in principle 
measurable. In this paper, we critically examine this claim and demonstrate that the phase obtained 
through this approach is actually gauge-dependent and not an observable for a nonclosed path. 
We also provide a brief critical discussion of the proposed experiment for observing the AB phase shift of a 
nonclosed path.
\end{abstract}

\maketitle

\section{Introduction} 
\label{sec_introduction}
The Aharonov-Bohm (AB) effect \cite{ehrenberg1949, aharonov1959}  is a quantum mechanical phenomenon in which a 
charged particle is influenced by the electromagnetic potential even in regions where the electromagnetic field 
can be neglected.  For reviews, see, e.g., \cite{peshkin1981, olariu1985, peshkin1989, wakamatsu2018}. 
The existence of the AB effect was first experimentally demonstrated by Tonomura {\it et al.} using electron beam holography 
techniques \cite{tonomura1986, osakabe1986}. 

In a typical scenario of the AB effect, we consider an external current source, such as a solenoid, and assume 
that a charged particle is moving in a region where the magnetic field generated by the current source is 
negligible, but the corresponding vector potential is finite. 
In the standard interpretation of the AB effect, the Schroedinger equation for the charged particle includes 
a coupling term between the particle and the background vector potential generated by the external current. 
As a result, the charged particle acquires an extra phase  even in regions where 
no magnetic field is present. 
This phase (the AB phase) is represented by a line integral of the background vector potential along the path 
of the particle.  Therefore it is widely accepted that the AB phase is a gauge-dependent quantity and is not 
an observable unless the particle's path forms a closed loop.  

However, there are some interesting unconventional approaches to understand the AB phase. 
For instance, Vaidman made an attempt to explain the AB phase without relying on gauge-dependent 
potentials  \cite{vaidman2012}. 
According to this explanation, the AB phase is attributed to the local interaction between the charged 
particle's field and the potential source. 
For further discussions on this approach, 
refer to \cite{aharonov2015,pearle2017,pearle2017-2, li2022}. 

There have been yet other proposals of new attempts based on a quantum electrodynamics 
framework \cite{santos1999,marletto2020,saldanha2021,kang2022}. 
In this approach, it is considered that a charged particle and an external current source interact 
by the exchange of quantum electrodynamics photons, with the AB phase being directly proportional 
to the change in interaction energy. 
Since energy is commonly regarded as a gauge-invariant observable, it is asserted that the AB phase 
shift is also gauge invariant and can be measured even when the particle's path is not closed 
\cite{ marletto2020, saldanha2021, kang2022}.  
In this paper, however,  we will disprove this claim by showing that the energy correction 
computed in the quantum electrodynamics framework generally depends on the choice of gauge. 

In Sec.~II, we first review the approach to the AB phase in the quantum electrodynamics framework. 
This approach primarily involves perturbations in both the charge of the particle, denoted as $e$, 
and the strength parameter of the external current, denoted as $g$.  
Next, we introduce a different scheme using the coherent state, where the assumption of $g$ being small 
is no longer required. In this scheme, while the results remain unchanged, the calculations are simplified, 
providing us with a deeper understanding of how the energy correction depends on the gauge condition. 
The Coulomb gauge or the Lorenz gauge is commonly used 
in the quantum electrodynamics approach to the AB phase.  
In Sec.~III, we adopt the axial gauge condition $A_3=0$, and demonstrate that 
the energy correction is a gauge dependent quantity, implying that the AB phase for a nonclosed 
path is not an observable. 
Discussions including brief comments on the proposed experiment for measuring the AB phase 
for a nonclosed path are given in Sec. IV.
%
\section{The system of a charged particle, quantum electromagnetic field, and an external current source} 
\label{sec_system} 
We consider the combined system of a charged particle, quantum electromagnetic field, and a static 
external current source such as a solenoid.  
The mass, the charge and the coordinate of the particle are 
denoted as $m$, $e$, and \mbold{q}, respectively. 
The charged particle interacting with the electromagnetic
field is also subject to the action of a certain potential $V(\mbold{q})$. 
The quantum electromagnetic field also couples to the current 
$J_\mu(\mbold{x}) = (0, \mbold{J}(\mbold{x}))$ produced by the static external current source. 
The equations of motion for the charged particle and the electromagnetic
fields can be written as follows:
\begin{subequations}
\label{eq_equations_of_motion}
\begin{align}
  &  m \frac{d^2 \mbold{q}}{dt^2} = -\frac{\del V}{\del \mbold{q}}+ 
             e \left( \mbold{E}(\mbold{q},t) 
              + \mbold{\dot{q}} \times \mbold{B}(\mbold{q},t) 
                 \right), 
               \\
 & \nabla\cdot\mbold{E}(\mbold{x},t)  = j^0(\mbold{x},t) , 
               \\
 &  \nabla \times \mbold{B}(\mbold{x},t)  = 
           \mbold{j}(\mbold{x},t) + g\mbold{J}(\mbold{x}) 
           +\frac{\del \mbold{E}(\mbold{x},t) }{\del t}, 
\end{align}
\end{subequations}
where $j^\mu(x)$ is the current of the charged particle given by 
\begin{align}
 & j^0(\mbold{x},t) =  e \delta(\mbold{x}-\mbold{q}), 
           \\
 & \mbold{j}(\mbold{x},t) =  
           e\dot{\mbold{q}}\delta(\mbold{x}-\mbold{q}), 
\end{align}
and we introduced the strength parameter $g$ for the external current $\mbold{J}(\mbold{x}) $ 
for later convenience. Note that the static external current 
$\mbold{J}(\mbold{x})$ is conserved: $\nabla \cdot \mbold{J}=0$. 
Furthermore, we take into account that the external current is localized in space. As a result, we can safely assume that all the fields rapidly approach zero at infinity.
Throughout this paper we employ the Heaviside-Lorentz (rationalized Gaussian) 
system of units with $\hbar = c = 1$. 

We start by adopting the Coulomb gauge, where 
the vector potential $\mbold{A}$ satisfies the transverse condition $\nabla \cdot \mbold{A} =0$, and 
for clarity the vector potential with this condition is denoted by $\mbold{A}_\perp$. 
The Hamiltonian of this system is then given by 
\begin{align}
 H_C =& \frac{1}{2m}(\mbold{p}-e\mbold{A}_\perp (\mbold{q}))^2 + V(\mbold{q}) 
                              \nonumber \\
       &+ \mint d^3x  \frac{\mbold{E}_\perp^2+\mbold{B}^2}{2}
      -g \mint d^3x \mbold{J}(\mbold{x}) \cdot \mbold{A}_\perp (\mbold{x}), 
      \label{eq_HC}
\end{align}
where $\mbold{E}_\perp$ is the transverse component of the electric field $\mbold{E}$, 
and the longitudinal component $\mbold{E}_\parallel$ and the time component 
$A_0$ are given by 
\begin{align}
  \mbold{E}_\parallel = -\nabla A_0,\ \ 
  A_0(\mbold{x},t) = \frac{1}{4\pi}\frac{e}{|\mbold{x}-\mbold{q}|}. 
\end{align}
We have the following equal-time commutation relations:   
\begin{align}
 &  [q^i,p^j]=i\delta^{ij} , \\
 &  [A_i^\perp (\mbold{x},t), E_j^\perp(\mbold{x}',t)] = 
     -i \delta_{ij}^{tr}(\mbold{x}-\mbold{x}'), 
\end{align}
for $i,j=1,2,3$, and 
all the other commutation relations vanish. 
Here $\delta_{ij}^{tr}(\mbold{x}-\mbold{x}')$ is the transverse
delta function defined by
\begin{align}
   \delta_{ij}^{tr}(\mbold{x}-\mbold{x}') = \mint \frac{d^3k}{(2\pi)^3} 
       e^{i\sbold{k}\cdot (\sbold{x}-\sbold{x}') }  
         \left( \delta_{ij}-\frac{k_i k_j}{\mbold{k}^2} \right). 
\end{align}
The fields $\mbold{A}_\perp$ and $\mbold{E}_\perp$ are expanded in terms of  
the photon creation and annihilation operators as 
\begin{align}
 \mbold{A}_\perp(\mbold{x}) =& \mint \frac{d^3k}{\sqrt{(2\pi)^3 2\omega}} 
                \sum_{\lambda=1}^2 \mbold{e}(k,\lambda)
                \nonumber \\
    &   \times \left( a(k,\lambda)e^{i\sbold{k}\cdot\sbold{x}}
          + a^\dagger(k,\lambda)e^{-i\sbold{k}\cdot\sbold{x}} \right) ,
                  \label{eq_Aexp}       \\
 \mbold{E}_\perp(\mbold{x}) =& i \mint
      \frac{d^3k}{\sqrt{(2\pi)^3}}\sqrt{\frac{\omega}{2}} \sum_{\lambda=1}^2 
          \mbold{e}(k,\lambda) 
             \nonumber \\
   & \times \left( a(k,\lambda)e^{i\sbold{k}\cdot\sbold{x}}
          - a^\dagger(k,\lambda)e^{-i\sbold{k}\cdot\sbold{x}} \right) , 
                \label{eq_Eexp}
\end{align}
where $\omega=|\mbold{k}|$ and $\mbold{e}(k,\lambda)$ is the transverse polarization 
vector satisfying $ \mbold{e}(k,\lambda) \cdot \mbold{k} =0$ for $\lambda=1,2$. 

\subsection{The effective Hamiltonian for the charged particle: The second order in 
$g$ and $e$}
\label{subsec_Saldanha}
The system described by the Hamiltonian of Eq.~(\ref{eq_HC}) consists of the charged 
particle and the electromagnetic fields.  To obtain an effective Hamiltonian for the 
charged particle, we need to eliminate the degrees of freedom of the electromagnetic fields.  
In this subsection we mainly follow the derivation by Saldanha \cite{saldanha2021}.  
See also \cite{santos1999,marletto2020,kang2022}.  

We write the Hamiltonian $H_C$ as 
\begin{align}
  H_C =& \frac{\mbold{p}^2}{2m} + V(\mbold{q}) + H_C',  
                  \label{eq_HCsplit}     \\
  H_C' =& H_{EM} + H_g + H_e + O(e^2), 
                 \label{eq_HCprime}
\end{align}
where 
\begin{align*}
 & H_{EM} = \mint d^3x  \frac{\mbold{E}_\perp^2+\mbold{B}^2}{2} 
                 = \sum_{\lambda=1}^2 \mint d^3k  \omega a^\dagger(k,\lambda) a(k,\lambda) , 
                          \\
  & H_g = -g \mint d^3x \mbold{J}(\mbold{x}) \cdot \mbold{A}_\perp(\mbold{x}),\ \ 
     H_e =  -e\frac{\mbold{p}}{m} \cdot \mbold{A}_\perp(\mbold{q}). 
\end{align*}
In this subsection we assume that the parameters $e$ and $g$ are small, and 
we calculate the energy of $H_C'$ up to the second order in $e$ and $g$. 
The resultant energy depends on the dynamical variables of the 
charged particle, $\mbold{q}$ and $\mbold{p}$, and it contributes part of the effective 
Hamiltonian of the charged particle.  

The unperturbed  state is the photon vacuum $\ket{0}$ that is the ground state of $H_{EM}$.  
It is evident that the first order energy correction vanishes, as the expectation value of $\mbold{A}$ in 
the vacuum $\ket{0}$ is 0.  The second order energy in parameters $e$ and $g$ is given by 
\begin{align}
   \bra{0} (H_e + H_g) \frac{Q}{E_0-H_{EM}} (H_e + H_g) \ket{0},  
        \label{eq_2nd}
\end{align}
where $Q=\mbold{1}-\ket{0}\bra{0}$ and $E_0$ is the unperturbed energy of the vacuum $\ket{0}$.  
Note that the terms of $O(e^2)$ and $O(g^2)$ are constants; they are independent of 
the dynamical variables of the charged particle. The relevant energy is, therefore, of $O(ge)$ 
and given by 
\begin{subequations}
\begin{align}
  \Delta \epsilon =&  \bra{0} H_g \frac{Q}{E_0-H_{EM}}H_e \ket{0}
    + {\rm c.c.}
         \label{eq_DeltaE_intforma}           \\
              =& \mint d^3k \sum_{\lambda=1}^2 \braket{0| H_g |k,\lambda}\frac{-1}{\omega} 
                                                                         \braket{k,\lambda | H_e | 0}  + {\rm c.c. }, 
         \label{eq_DeltaE_intformb}
\end{align}
          \label{eq_DeltaE_intform}
\end{subequations}
with c.c. representing the complex conjugate of the terms preceding it. 
The intermediate state $\ket{k,\lambda}$ is the one-photon state of momentum $k$ and 
polarization $\lambda=1,2$. 
Using the Fourier expansion of $\mbold{A}_\perp$ of Eq.~(\ref{eq_Aexp}), we have 
\begin{align}
  H_g =& - g\sum_{\lambda=1}^2 \mint d^3k \frac{1}{\sqrt{2\omega}} 
   \mbold{e}(k,\lambda) 
   \nonumber \\
   & \cdot 
   \left(
        a(k,\lambda) \mbold{J}_{\sbold{k}}^*
      + a^\dagger(k,\lambda) \mbold{J}_{\sbold{k}}) 
   \right), 
\end{align}
where $\mbold{J}_{\sbold{k}}$ is the Fourier transform of the external current 
source $\mbold{J}(\mbold{x})$:  
\begin{align}
  \mbold{J}_{\sbold{k}} = \frac{1}{\sqrt{(2\pi)^3}} \mint d^3x 
     \mbold{J}(\mbold{x}) e^{-i\sbold{k}\cdot\sbold{x}}. 
\end{align}
We can now calculate $\Delta \epsilon$ of Eq.~(\ref{eq_DeltaE_intform}).  We  find  
\begin{align}
  \Delta \epsilon = -eg \frac{\mbold{p}}{m} \cdot \mbold{A}^\perp_{ext}(\mbold{q}) , 
                \label{eq_Delta_E} 
\end{align} 
where $\mbold{A}^\perp_{ext}(\mbold{x})$ is defined to be 
\begin{align}   
    \mbold{A}_{ext}^\perp (\mbold{x}) =& \frac{1}{\sqrt{(2\pi)^3}} \mint d^3k 
     \frac{\mbold{J}_{\sbold{k}}}{\mbold{k}^2} e^{i\sbold{k}\cdot\sbold{x}}
                        \nonumber \\
       =& \frac{1}{4\pi} \mint d^3x' 
      \frac{\mbold{J}(\mbold{x}')}{|\mbold{x}-\mbold{x}'|} , 
    \label{eq_Aext_perp}
\end{align}
which is just the vector potential generated by the current source 
$\mbold{J}(\mbold{x})$ according to the Biot-Savart law.    
This $\mbold{A}_{ext}^\perp(\mbold{x})$ satisfies the transverse condition 
$\nabla \cdot \mbold{A}_{ext}^\perp(\mbold{x}) = 0$, which follows from 
the current conservation law $\nabla \cdot \mbold{J}(\mbold{x}) =0$. 

Thus the effective Hamiltonian for the charged particle is given by 
\begin{align}
  h_C = \frac{\mbold{p}^2}{2m} + V(\mbold{q})  -eg \frac{\mbold{p}}{m} \cdot \mbold{A}^\perp_{ext}(\mbold{q}) 
                + O(e^2). 
      \label{eq_hC}
\end{align} 
Note that the $h_C$ takes the gauge covariant form 
\begin{align}
  h_C = \frac{( \mbold{p} -eg \mbold{A}^\perp_{ext}(\mbold{q}))^2}{2m} + V(\mbold{q}) , 
     \label{eq_hC_full}
\end{align}
if the appropriate higher order term of $O(g^2e^2)$ is added. 
This can be interpreted as representing the Hamiltonian of a charged particle moving 
under the influence of the external potential $\mbold{A}^\perp_{ext}(\mbold{q})$.  

In this approach, the coupling term in the effective Hamiltonian given by Eq.~(\ref{eq_Delta_E}) is the second-order 
energy correction.  
Suppose that the charged particle moves along a path $L$: from $t=t_1$ to $t=t_2$.  
The AB phase shift acquired by the particle is then given by a time integral of $\Delta \epsilon$: 
\begin{align*}
 \Phi_{AB} = -eg \int_{t_1}^{t_2}\!\! dt \frac{\mbold{p}}{m} \cdot \mbold{A}_{ext}^\perp (\mbold{q})
    = -eg \int_{L}\!\! d\mbold{q} \cdot \mbold{A}_{ext}^\perp (\mbold{q}), 
\end{align*}
while energy is generally considered to be a gauge-invariant observable.  
It is this observation that underlies the claim that 
the AB phase shift is a gauge invariant and measurable observable even when the path $L$ is not closed 
\cite{ marletto2020, saldanha2021, kang2022}.  

In the subsequent sections, we will refute this assertion by showing that the energy correction 
$\Delta \epsilon$ is generally gauge dependent. 
In the Lorenz gauge, the form of $\Delta \epsilon$ given in Eq.~(\ref{eq_DeltaE_intforma})  
remains unaltered, although the intermediate states should include scalar and longitudinal photon 
states alongside the transverse ones. However, it is not difficult to see that  
the energy correction $\Delta \epsilon$ coincides with that in the Coulomb gauge .  
In Sec.~\ref{sec_gaugedependence}, we will choose the axial gauge to demonstrate the 
energy correction $\Delta \epsilon$ is a gauge dependent quantity.  In the next subsection, however, we 
present the new scheme using the coherent state where the parameter $g$ is not assumed to be small. 
We will see that this scheme gives the same result as the one presented here 
but provides more insight into how the energy correction depends on the gauge condition. 

\subsection{Coherent state scheme: The first order in $e$}
\label{subsec_coherent} 
In this subsection, we present the method that uses photon coherent states without assuming that the parameter $g$ is small. 
The results we will obtain are the same as those in the preceding subsection. 
However, the calculations are simplified in this method. It also provides  better insight into the gauge dependence 
of the energy correction in question. 
In this method we treat $H_C'$ of Eq.~(\ref{eq_HCprime}) 
in the fist-order perturbation in $e$.  The unperturbed Hamiltonian is then $H_{EM}+H_g$, which is expressed as 
\begin{align*}
& H_{EM}+H_g 
        \\
 =& \sum_{\lambda=1}^2 \mint d^3k  \omega a^\dagger(k,\lambda) a(k,\lambda) 
         \\
    & - g\sum_{\lambda=1}^2 \mint  \frac{d^3k}{\sqrt{2\omega}} \mbold{e}(k,\lambda) 
    \cdot 
   \left(
        a(k,\lambda) \mbold{J}_{\sbold{k}}^*
      + a^\dagger(k,\lambda) \mbold{J}_{\sbold{k}}
   \right).
\end{align*} 
This Hamiltonian can be ``diagonalized''  in terms of the  photon annihilation and creation operators, 
$\tilde a(k,\lambda)$ and $\tilde a^\dagger(k,\lambda)$, which are defined through 
\begin{align}
   a(k,\lambda) = \tilde a(k,\lambda) + \alpha(k,\lambda), 
\end{align}
with $\alpha(k,\lambda) = g/\sqrt{2\omega^3} \mbold{e}(k,\lambda) \cdot \mbold{J}_{\sbold{k}}$. 
Up to a constant the result is given by 
\begin{align}
  H_{EM}+ H_g  = 
    \sum_{\lambda=1}^2 \int d^3k\, \omega\, 
           \tilde a^\dagger(k,\lambda) \tilde a(k,\lambda). 
    \label{eq_HEM_Hg}
\end{align} 
The ground state of $H_{EM}+ H_g$, denoted by $\ket{\tilde 0}$, is then the 
state that is annihilated by $\tilde a(k,\lambda)$; that is 
$\tilde a(k,\lambda) \ket{\tilde 0}=0$. The state $\ket{\tilde 0}$ is  therefore an eigenstate of the annihilation operator $a(k,\lambda)$, 
the coherent state of photons \cite{gerry2005}. 
\begin{align}
   a(k,\lambda) \ket{\tilde 0} = \alpha(k,\lambda) \ket{\tilde 0}. 
\end{align}
Note that the expectation value of $\mbold{A}_\perp$ with respect to 
the coherent state $\ket{\tilde 0}$ is not 0, but given by $g$ times $\mbold{A}_{ext}^\perp(\mbold{x})$ 
of Eq.~(\ref{eq_Aext_perp}). 
\begin{align}
 \braket{\tilde 0 | \mbold{A}_\perp(\mbold{x}) | \tilde 0} =& \mint \frac{d^3k}{\sqrt{(2\pi)^3 2\omega}} 
                \sum_{\lambda=1}^2 \mbold{e}(k,\lambda)
                \nonumber \\
    &   \times \left( \alpha(k,\lambda)e^{i\sbold{k}\cdot\sbold{x}}
          + \alpha^\dagger(k,\lambda)e^{-i\sbold{k}\cdot\sbold{x}} \right) 
                \nonumber \\
  =&  g \frac{1}{4\pi} \mint d^3x' 
      \frac{\mbold{J}(\mbold{x}')}{|\mbold{x}-\mbold{x}'|}         
      = g \mbold{A}_{ext}^\perp(\mbold{x}). 
                  \label{eq_exp_of_A}
\end{align}
The unperturbed energy $\tilde E_0$, which is a constant, can be taken to be 0.  
The first order energy correction in $e$ is simply given by the expectation value of $H_e$ in 
the unperturbed state $\ket{\tilde 0}$: 
\begin{align}
 \Delta E =&
  \braket{\tilde 0 | H_e | \tilde 0}=    
    -e\frac{\mbold{p}}{m} \cdot  \braket{\tilde 0 |\mbold{A}_\perp(\mbold{q})  | \tilde 0} 
                    \nonumber \\
    =& -eg\frac{\mbold{p}}{m}  \mbold{A}_{ext}^\perp(\mbold{q}). 
      \label{eq_coupling}
\end{align}
This is exactly equal to the energy correction $\Delta \epsilon$ of Eq.~(\ref{eq_Delta_E}) that is the result of 
the second order perturbation in $e$ and $g$. Thus we have obtained 
the same effective particle Hamiltonian $h_C$ as that given in Eq.~(\ref{eq_hC}).  

A few remarks are now in order. 
First, the results obtained in the preceding subsection \ref{subsec_Saldanha} hold true regardless of the magnitude of $g$. 
This is not a coincidence, but is due to the following reasons: Equation (\ref{eq_exp_of_A}) shows that 
the expectation value $\braket{\tilde 0 | \mbold{A}_\perp(\mbold{x}) | \tilde 0}$  is linear in $g$ though 
the coherent state $\ket{\tilde 0}$ itself contains higher-order terms in $g$.  
Suppose that we calculate the expectation value of $ \mbold{A}_\perp(\mbold{x})$ in the state $\ket{0_1}$ that is 
the approximate ground state of $H_{EM}+H_g$ in the first-order perturbation in $g$. 
\begin{align*}
  \ket{0_1} = \ket{0} + 
   \frac{Q}{E_0 - H_{EM}}H_g \ket{0}.
\end{align*}
We then have 
\begin{align*}
  \braket{\tilde 0 |  \mbold{A}_\perp(\mbold{x}) | \tilde 0} =&  \braket{ 0_1 |  \mbold{A}_\perp(\mbold{x}) | 0_1}+ \delta
                                        \nonumber \\
  =& \bra{0} H_g \frac{Q}{E_0-H_{EM}} \mbold{A}_\perp(\mbold{x}) \ket{0} +{\rm c.c.} + \delta,
\end{align*}
where the error $\delta$ is  $O(g^2)$. 
Notice that the left-hand side is order $O(g)$. This implies that $\delta$ should be 
exactly 0 as the terms other than $\delta$ on the right-hand side are order $O(g)$ . 
Thus we find 
\begin{align}
  \braket{\tilde 0 | H_e | \tilde 0} =
   \bra{0} H_g \frac{Q}{E_0-H_{EM}}H_e \ket{0} +{\rm c.c.}, 
\end{align}
which is just the second-order energy correction of Eq.~(\ref{eq_DeltaE_intform}) 
obtained in the preceding subsection. 

Second, as can be seen in Eq.~(\ref{eq_coupling}), the coupling term in the effective Hamiltonian, 
which is identified with the energy correction $\Delta E = \Delta \epsilon$, is expressed in terms of 
the expectation value of the vector potential in the coherent state. 
This expectation value should reflect the gauge condition imposed on the vector potential. 
This strongly suggests that the coupling term depends on the gauge that we choose. 
We will examine specifically the case of the axial gauge in Sec.~\ref{sec_gaugedependence}. 

Third, this concerns the change in magnetic field energy due to the motion of the charged particle.
Using classical electromagnetism, Boyer calculated this quantity and obtained the following result \cite{boyer1971}: 
\begin{align}
   \Delta {\cal E}_{\rm Boyer} = eg\frac{\mbold{p}}{m}  \mbold{A}_{ext}^\perp(\mbold{q}), 
              \label{eq_boyer}
\end{align}
which has the same form as the coupling term between the charged 
particle and the external potential $\mbold{A}_{ext}^\perp(\mbold{q})$. 
However, no attention seems to have been paid to the fact that the signs are different.
Where does the Boyer's energy (\ref{eq_boyer}) appear in our treatment, in which 
the energy correction of $O(e)$ is  given by $\braket{\tilde 0 | H_e | \tilde 0}$ only? 
The answer is that it is hidden as part of the expectation value of the unperturbed Hamiltonian. 
Let $\ket{\tilde \phi_1}$ be the first-order perturbed eigenstate of $H_{EM}+H_g+H_e$. 
\begin{align}
  \ket{\tilde \phi_1} =
    \ket{\tilde 0} + \frac{\tilde Q}{\tilde E_0-(H_{EM}+H_g)}H_e \ket{\tilde 0},  
         \label{eq_tilde_phi_1}
\end{align}
with $\tilde Q = \mbold{1}-\ket{\tilde 0}\bra{\tilde 0}$. 
Now consider the expectation value of the unperturbed Hamiltonian in the state $\ket{\tilde \phi_1}$. 
The result should be given  by 
\begin{align}
  \braket{\tilde \phi_1 | H_{EM}+ H_g  | \tilde \phi_1} = \tilde E_0 + O(e^2) , 
\end{align}
with no terms of order $O(e)$.  This, however, does not necessarily imply that neither 
$\braket{\tilde \phi_1 | H_{EM}| \tilde \phi_1}$ nor $\braket{\tilde \phi_1 | H_g  | \tilde \phi_1}$ 
contains a contribution of order $O(e)$.  
It turns out that $\braket{\tilde \phi_1 | H_{EM}| \tilde \phi_1}$ is given by 
$\Delta {\cal E}_{\rm Boyer}$ up to a constant, but is  canceled by the 
contribution from $\braket{\tilde \phi_1 | H_g  | \tilde \phi_1}$.  
See the Appendix~\ref{app_Boyer} for details. 

\section{The AB-phase of a nonclosed path is not observable}
\label{sec_gaugedependence} 
In this section we show that the coupling term 
in the effective Hamiltonian depends on the gauge. 
To do so we treat the same system as that discussed in the preceding section by 
imposing the axial gauge condition, $A_3=0$. 
Note that this condition completely fixes the vector 
potential as we assume all the fields should approach zero at infinity. 
For canonical quantization of electromagnetic fields in the axial gauge, we refer the reader to  
Ref.~\cite{haller1994}. 
The Hamiltonian is given by 
\begin{align}
 H_X =& \frac{1}{2m}(\mbold{p}-e\mbold{A}(\mbold{q}))^2 + V(\mbold{q}) 
       + \mint d^3x  \frac{\mbold{E}^2+\mbold{B}^2}{2}
                             \nonumber \\
                   & -g\mint d^3x \mbold{J}(\mbold{x}) \cdot \mbold{A}(\mbold{x}).  
                   \label{eq_Haxial_1}
\end{align}
Here, for $\mbold{q}, \mbold{p}$,  and the $x$- and $y$- components of 
$\mbold{A}$ and $\mbold{E}$,  the following typical  canonical commutation relations 
are assumed:
\begin{align*}
 & [q^k,q^l] = [p^k,p,^l]=0,\ [q^k,p^l]=i\delta_{kl},  \\
 & [A_i(\mbold{x}), A_j(\mbold{x}')] = [E_i(\mbold{x}), E_j(\mbold{x}')] 
   = 0,  \\
 & [A_i(\mbold{x}), E_j(\mbold{x}')] 
    = -i \delta_{ij}\delta(\mbold{x}-\mbold{x}'),  \\
 &  [q^k,A_i(\mbold{x})] = [q^k,E_i(\mbold{x})] = 
 [p^k,A_i(\mbold{x})] = [p^k,E_i(\mbold{x})] = 0,  
\end{align*}
for $k,l=1,2,3$ and $i,j=1,2$. 
The other commutation relations are determined by the constraint conditions 
in the axial gauge.  
First, we have the gauge fixing condition:  $A_3(\mbold{x}) = 0$. 
Second, the $z$ component of $\mbold{E}$ is constrained to be 
\begin{align}
  E_3(\mbold{x}) = -\frac{1}{\del_3} 
    \left( \sum_{i=1}^2 \del_i E_i(\mbold{x}) + j_0(\mbold{x}) \right) , 
        \label{eq_const_divE}
\end{align}
so that Gauss's law $\nabla \cdot \mbold{E} = j_0$ is fulfilled. 
Third, the component $A_0$ is given by $A_0(\mbold{x}) = \del_3^{-1} E_3(\mbold{x})$ , 
 which respects the relations $E_3 = \del_3 A_0 - \del_0 A_3$ with $A_3=0$.  
Under these commutation relations and constraints , one can verify that 
the Hamiltonian $H_X$ correctly reproduces the equations of motion of Eq.~(\ref{eq_equations_of_motion}). 

The vector potential $\mbold{A}(\mbold{x})$ and the electric field $\mbold{E}(\mbold{x})$ 
in the axial gauge are related to 
the transverse vector potential $\mbold{A}^\perp(\mbold{x})$  and  
the transverse electric field $\mbold{E}^\perp(\mbold{x})$, respectively: 
\begin{align}
 A_i(\mbold{x}) =& A_i^\perp(\mbold{x}) - \frac{\del_i}{\del_3}A_3^\perp(\mbold{x}) 
     \ \ \ i=1,2,3, 
                           \\
 E_i(\mbold{x}) =& E_i^\perp(\mbold{x}) - \delta_{i3}\frac{1}{\del_3} j_0(\mbold{x})
     \ \ \ i=1,2,3, 
\end{align}
with the Fourier expansions of $\mbold{A}^\perp(\mbold{x})$ and 
$\mbold{E}^\perp(\mbold{x})$ given in Eqs.~(\ref{eq_Aexp}) and (\ref{eq_Eexp}), 
respectively. 
Using these Fourier expansions, one can verify that the three aforementioned 
constraints in the axial gauge are satisfied and all the related commutation relations follow.  
The Hamiltonian of Eq.~(\ref{eq_Haxial_1})  can be written as 
\begin{align*}
 H_X =& \frac{1}{2m}(\mbold{p}-e\mbold{A}(\mbold{q}))^2 
             + V(\mbold{q}) 
               \nonumber  \\
            &  + \mint d^3k\, \omega \sum_{\lambda=1}^2 
                                               a^\dagger(k,\lambda) a(k,\lambda)
           \nonumber \\
  &-e\frac{1}{(\del_3)^2} \left( \sum_{i=1}^2 \del_i E_i(\mbold{q}) \right)
   -g\mint d^3x \mbold{J}(\mbold{x}) \cdot \mbold{A}(\mbold{x}). 
\end{align*}

We are now ready to calculate the effective Hamiltonian for the charged particle in the axial gauge.  
As in the case of the Coulomb gauge, we write 
\begin{align}
  H_X =& \frac{\mbold{p}^2}{2m} + V(\mbold{q}) + H_X',  
                 \label{eq_HXsplit}     \\
  H_X' =& H_{EM} + H_g + H_e^{(1)} +H_e^{(2)} + O(e^2), 
                 \label{eq_HXprime}
\end{align}
where 
\begin{align*}
 & H_{EM} = \sum_{\lambda=1}^2 \mint d^3k  \omega a^\dagger(k,\lambda) a(k,\lambda) , 
                          \\
  & H_g = -g \mint d^3x \mbold{J}(\mbold{x}) \cdot \mbold{A}(\mbold{x}),
                          \\
  & H_e^{(1)} =  -e\frac{\mbold{p}}{m} \cdot \mbold{A}(\mbold{q}), 
                          \\
  & H_e^{(2)} = -e\frac{1}{(\del_3)^2}  \left( \sum_{i=1}^2 \del_i E_i(\mbold{q}) \right). 
\end{align*}

We evaluate $H_X'$ in terms of the first-order perturbation theory of $e$, 
as we did for the Coulomb gauge in Subsec.~\ref{subsec_coherent}.  
The Hamiltonian $H_g$ is expanded as 
\begin{align}
  H_g =& - g\sum_{\lambda=1}^2 \mint d^3k \frac{1}{\sqrt{2\omega}} 
   \mbold{e}^X(k,\lambda) 
   \nonumber \\
   & \cdot 
   \left(
        a(k,\lambda) \mbold{J}_{\sbold{k}}^*
      + a^\dagger(k,\lambda) \mbold{J}_{\sbold{k}}) 
   \right), 
\end{align}
where
\begin{align}
  e^X_i(k,\lambda) = e_i(k,\lambda)-\frac{k_i}{k_3}e_3(k,\lambda)\ \ i=1,2,3. 
\end{align}
is the polarization vector of the vector potential $\mbold{A}(\mbold{x})$ in 
the axial gauge. In the above expression of $H_g$, however,  this polarization vector 
$\mbold{e}^X(k,\lambda)$ 
can be replaced by the transverse polarization vector $\mbold{e}(k,\lambda)$, as 
the current conservation of the external current, 
$\mbold{k}\cdot \mbold{J}_{\sbold{k}}=0$, implies 
$\mbold{e}^X(k,\lambda) \cdot \mbold{J}_{\sbold{k}}=
\mbold{e}(k,\lambda) \cdot \mbold{J}_{\sbold{k}}$. 
Thus the unperturbed Hamiltonian $H_{EM}+H_g$ is  the same as that 
, given by Eq.~(\ref{eq_HEM_Hg}), in the Coulomb gauge. The unperturbed ground state is therefore the coherent state 
$\ket{\tilde 0}$ defined through $a(k,\lambda) \ket{\tilde 0} = \alpha(k,\lambda) \ket{\tilde 0}$ 
with  $\alpha(k,\lambda) = g/\sqrt{2\omega^3} \mbold{e}(k,\lambda) \cdot \mbold{J}_{\sbold{k}}$.

The expectation value of $\mbold{A}(\mbold{x})$ in the coherent state $\ket{\tilde 0}$ can 
be calculated as
\begin{align}
    \braket{\tilde 0 | \mbold{A}(\mbold{x}) | \tilde 0} =& 
       \braket{\tilde 0 |   A^\perp(\mbold{x}) - \frac{\nabla}{\del_3}A_3^\perp(\mbold{x})  | \tilde 0} 
                                   \nonumber \\ 
   =& g\left( \mbold{A}^\perp_{ext}(\mbold{x}) - \frac{\nabla}{\del_3} A_{\perp,ext}^3(\mbold{x}) \right)
                                  \nonumber \\
   \equiv & g \mbold{A}^X_{ext}(\mbold{x}), 
\end{align}
where $\mbold{A}^\perp_{ext}(\mbold{x})$ is defined in Eq.~(\ref{eq_Aext_perp}). 
Remember that $\mbold{A}^\perp_{ext}(\mbold{x})$ is the transverse vector potential 
generated by the external current. 
 This implies that $\mbold{A}^X_{ext}(\mbold{x})$ is also the vector potential generated 
 by the external current, but with 
 the axial gauge condition $\mbold{A}^{X,3}_{ext}(\mbold{x})=0$.  As for the electric 
 field $\mbold{E}$, it can be readily checked that $\braket{\tilde 0 | E_i(\mbold{x}) | \tilde 0} = 0$ for $i=1,2$. 
 
The unperturbed energy is constant and taken to be 0, and 
the first-order energy correction in $e$ is  given by 
\begin{align}
 \Delta E = 
 \braket{\tilde 0 | H_e^{(1)}+H_e^{(2)} | \tilde 0} 
 = -eg\frac{\mbold{p}}{m} \cdot \mbold{A}^{X}_{ext}(\mbold{q}) ,  
       \label{eq_He_axial}
\end{align}
which leads to the particle effective Hamiltonian:  
\begin{align}
  h_X = \frac{1}{2m}\mbold{p}^2 + V(\mbold{q}) 
      -eg\frac{\mbold{p}}{m} \cdot \mbold{A}_{ext}^{X}(\mbold{q}) + O(e^2). 
\end{align}
In general the vector potentials $\mbold{A}_{ext}^\perp(\mbold{q})$ and  $\mbold{A}_{ext}^X(\mbold{q})$ 
are different for a given external current $\mbold{J}(\mbold{x})$, implying that 
the AB phase acquired by the charged particle along a nonclosed path depends on the gauge 
that we choose, and it is not an observable. 

We obtained the effective Hamiltonian $h_X$ by using the first-order perturbation theory of $e$, 
as we did in Subsec. \ref{subsec_coherent} for the case of Coulomb gauge. 
The result should be unchanged if we perform the second-order perturbation in $g$ and $e$ as 
the calculation described in Subsec. \ref{subsec_Saldanha}. 
Before concluding this subsection, we will confirm this equivalence.  

In the perturbation in $g$ and $e$, the unperturbed state is $\ket{0}$ given by 
the ground state of $H_{EM}$, and $H_g +H_e^{(1)} +H_e^{(2)}$ is 
the perturbation. Since the first-order energy correction is 0, we consider the 
energy correction of order $O(ge)$: 
\begin{align}
 \Delta \epsilon = 
  \braket{0| H_g \frac{1}{E_0-H_{EM}}(H_e^{(1)}+H_e^{(2)}) |0} + {\rm c.c.}
        \label{eq_Saldanha_Delta_E_axial}
\end{align}
As in the case of the Coulomb gauge, this $\Delta \epsilon$ can be expressed as 
\begin{align}
 \Delta \epsilon= \braket{0_1 | H_e^{(1)}+H_e^{(2)} |0_1} + O(g^2), 
     \label{eq_Saldanha_Delta_E_axial_He}
\end{align}
with $\ket{0_1}$ being the approximate ground state of $H_{EM}+H_g$ in the first-order 
perturbation in $g$. 
In this way, we can see that the calculation of perturbation in $g$ and $e$ will eventually lead to 
the result of Eq.~(\ref{eq_He_axial}), but let us now analyze 
$\Delta \epsilon$ in Eq.~(\ref{eq_Saldanha_Delta_E_axial}) explicitly.  
First, after some involved calculation, we find 
\begin{align}
  &  \braket{0| H_g \frac{1}{E_0-H_{EM}}H_e^{(1)}|0} 
                   \nonumber \\
 =& -\frac{eg}{2m} \mbold{p} \cdot \left( \mbold{A}_{ext}^{\perp}(\mbold{q}) -
          \frac{\nabla}{\del_3} A_{\perp,ext}^{3}(\mbold{q}) \right)
               \nonumber \\
 =& -\frac{eg}{2m} \mbold{p} \cdot \mbold{A}_{ext}^{X}(\mbold{q}). 
     \label{eq_Hg_He}
\end{align}
Second, the following matrix element turns out to be purely imaginary:
\begin{align}
 \braket{0| H_g \frac{1}{E_0-H_{EM}}H_e^{(2)}|0}. 
\end{align}
Combining these results, we conclude that $\Delta \epsilon$ is eventually given by 
$\Delta E$ of  Eq.~(\ref{eq_He_axial}), 
which is as expected from Eq.~(\ref{eq_Saldanha_Delta_E_axial_He}).  

\section{Discussion and conclusions}
\label{sec_discussion}
The purpose of this paper is to investigate whether the AB phase shift in the case of a nonclosed path is 
independent of the gauge and therefore measurable. In the recent approach 
\cite{santos1999, marletto2020, saldanha2021, kang2022}, the coupling term between 
the charged particle and the electromagnetic potential is identified with the energy change $\Delta E$ resulting from 
the quantum mechanical electromagnetic interaction between the charged particle and the external current 
source. Since energy is generally thought to be  gauge-invariant, this leads to the claim that 
the AB phase shift is a gauge invariant and measurable observable even for a nonclosed path 
\cite{marletto2020, saldanha2021,kang2022}.  
We have disproved  this claim by explicitly showing that the energy correction 
$\Delta E$ is generally gauge dependent. 

One may still wonder why the energy change $\Delta E$  can be gauge-dependent.  
Let $H_C$ and $H_X$ be the Hamiltonians in the Coulomb gauge and the axial gauge, respectively. 
As operators we know that  $H_C \ne H_X$, though we believe that their eigenvalues are the same. 
However, we do not try to directly determine the eigenvalues of $H_C$ or $H_X$.  
What we do in this approach is to eliminate the degrees of freedom of the electromagnetic field 
and obtain an effective Hamiltonian $h$ for the charged particle: 
\begin{align}
 h = \frac{1}{2m}\mbold{p}^2 +V(\mbold{q}) 
         - eg \frac{\mbold{p}}{m}\cdot 
             \braket{\tilde 0 |\mbold{A}(\mbold{q}) |\tilde 0}+ \cdots.
\end{align}
Starting from $H_C$ yields $h_C$, and starting from $H_X$ yields $h_X$. Generally, 
we have $h_C \ne h_X$, as $\mbold{A}_{ext}^\perp(\mbold{x}) \ne \mbold{A}^{X}_{ext}(\mbold{x})$. 
To calculate the energy of the system, one must still determine the eigenvalues of $h_C$ or $h_X$. 
They should be the same. In conclusion, it may be misleading to say 
that this approach "calculates the energy change of the system." 
It would be more appropriate to say that it "derives the effective Hamiltonian for the charged particle." 
From this perspective, it seems natural for the effective Hamiltonian to depend on the gauge choice.  

We have found that the AB phase for a nonclosed path cannot be a 
measurable physical quantity. On the other hand, some experiments 
have been proposed \cite{marletto2020,saldanha2021,kang2022} that 
appear to allow the measurement of the AB phase along a nonclosed path. 
This is a contradiction that necessitates an examination of the 
proposed experiments. 
Here we take the experiment proposed in \cite{saldanha2021} because it appears to be the simplest conceptually.
Essentially, it involves abruptly cutting off the current in a solenoid before the charged particle completes a closed path. 
The assumption here is that this would cause the charged particle to remember its phase change at the time of the interruption, 
and the interference pattern at the closed path would reflect the phase change at the intermediate point in the path. 
The problem here, however, is that a time-varying magnetic field produces an electric field. 
If $A_\mu(x)$ is time dependent, then the closed curve drawn by the charged particle would be a closed curve 
in four-dimensional space-time. 
In other words, when the solenoid current changes, the charged particle also 
undergoes a phase change with a contribution of $ \int_C A_0(x)dt$, and it does not retain the phase 
difference at the time when the solenoid current is suddenly cut. 
One way to demonstrate this would be to assume that the current $J_\mu (x)$ in the solenoid is time dependent 
and to show that the phase difference after drawing a closed curve in four-dimensional space-time is independent of the gauge, 
while the phase accumulated during the path is gauge dependent. 
For a more comprehensive discussion on the four-dimensional loop integral of the time-dependent vector potential, we refer 
the reader to Ref.~\cite{horvat2020}. 

\smallskip
{\it Note added:} Recently,  Wakamatsu \cite{wakamatsu2023}
showed that the AB phase depends on the residual gauge freedom in the 
Coulomb gauge within a quantum electrodynamics approach.


\begin{acknowledgments}
The author would like to thank M. Wakamatsu for numerous useful discussions and for critical comments on the manuscript. 
It is our pleasure to thank Sebastian Horvat for bringing Ref. \cite{horvat2020} to our attention.
\end{acknowledgments}

\appendix*
\smallskip
\section{The Boyer's energy change}
\label{app_Boyer}
We first rewrite $H_{EM}$ as 
\begin{align*}
  H_{EM} =& \mint d^3x  \frac{\mbold{E}_\perp^2+\mbold{B}^2}{2} 
                           \\
         =& \mint d^3x  \frac{\mbold{E}_\perp^2 + \tilde{\mbold{B}}^2}{2}
          + \mint d^3x \tilde{\mbold{B}} \cdot \mbold{B}_{ext} 
          + \text{const},
\end{align*}
where $\mbold{B}_{ext}= g \nabla \times \mbold{A}_{ext}^\perp$ 
and $\tilde{\mbold{B}} = \mbold{B}-\mbold{B}_{ext}$. 
Then we find that, up to a constant, 
\begin{align*}
  \braket{\tilde \phi_1 | H_{EM} | \tilde \phi_1} 
  = \mint d^3x \braket{\tilde \phi_1 | \tilde{\mbold{B}}| \tilde \phi_1}
              \cdot \mbold{B}_{ext} ,
\end{align*}
with $\ket{\tilde \phi_1}$ defined in Eq.~(\ref{eq_tilde_phi_1}).  
After some calculation, we find that 
\begin{align}
  \braket{\tilde \phi_1 | \tilde{\mbold{B}}(\mbold{x})| \tilde \phi_1}
 = -\frac{\mbold{p}}{m} \times \nabla \frac{1}{4\pi} 
      \frac{e}{|\mbold{x}-\mbold{q}|} + O(e^2), 
\end{align}
which is the nonrelativistic expression of the magnetic field generated by 
the charged particle moving at velocity $\mbold{p}/m$. 

Using this expression, we have 
\begin{align*}
 & \braket{\tilde \phi_1 | H_{EM} | \tilde \phi_1}
    = -\frac{e}{4\pi m}  \mint d^3x \left( \mbold{p} \times 
         \nabla \frac{1}{|\mbold{x}-\mbold{q}|} \right) 
          \cdot \mbold{B}_{ext}(\mbold{x}) 
        \nonumber \\
 &= \frac{e}{4\pi m} \mbold{p} \cdot 
       \mint d^3x \frac{1}{|\mbold{x}-\mbold{q}|} 
          \nabla \times  \mbold{B}_{ext}(\mbold{x}) 
       \nonumber \\
 &= \frac{eg}{4\pi m} \mbold{p} \cdot 
       \mint d^3x \frac{\mbold{J}(\mbold{x})}{|\mbold{x}-\mbold{q}|} 
  = eg\frac{\mbold{p}}{m} \cdot \mbold{A}_{ext}^\perp(\mbold{q}), 
\end{align*}
where $\mbold{A}_{ext}^\perp$ is defined in Eq.~(\ref{eq_Aext_perp}).  
This is the Boyer's energy change $\Delta {\cal E}_{\rm Boyer}$ given in Eq.~(\ref{eq_boyer}). 
We now observe that $H_g$ can be rewritten as follows: 
\begin{align*}
 H_g =& -g \mint d^3x \mbold{J}\cdot \mbold{A_\perp} 
     =  - \mint d^3x  \nabla \times  \mbold{B}_{ext}  \cdot \mbold{A}^\perp
                      \\
     =&  -\mint d^3x  \mbold{B}_{ext}  \cdot \nabla \times \mbold{A}^\perp
     = -\mint d^3x  \mbold{B}_{ext}  \cdot \tilde{\mbold{B}} + {\rm const}. 
\end{align*}
Therefore, up to a constant, we conclude 
\begin{align}
  \braket{\tilde \phi_1 | H_g | \tilde \phi_1} = 
    -eg\frac{\mbold{p}}{m} \cdot \mbold{A}_{ext}^\perp(\mbold{q}), 
\end{align}
which cancels the Boyer's energy $\braket{\tilde \phi_1 | H_{EM} | \tilde \phi_1}$. 



\end{document}